# CONSTRUCTING PATH EFFICIENT AND ENERGY AWARE VIRTUAL MULTICAST BACKBONES IN STATIC AD HOC WIRELESS NETWORKS

Tamaghna Acharya[1], Samiran Chattopadhyay[2] and Rajarshi Roy[3]

[1]Dept. of ETCE, BESU, Shibpur, Howrah 711103, India
t_acharya@telecom.becs.ac.in
[2]Dept. of Information Technology, Jadavpur University, Kolkata-700098, India
samiranc@it.jusl.ac.in
[3]Dept. of Electronics and Electrical Communication Engineering,
Indian Institute of Technology, Kharagpur, India
royr@ece.iitkgp.ernet.in

## ABSTRACT

For stationary wireless ad hoc networks, one of the key challenging issues in routing and multicasting is to conserve as much energy as possible without compromising path efficiency measured as end-to-end delay. In this paper, we address the problem of path efficient and energy aware multicasting in static wireless ad hoc networks. We propose a novel distributed scalable algorithm for finding a virtual multicast backbone (VMB). Based on this VMB, we have further developed a multicasting scheme that jointly improves path efficiency and energy conservation. By exploiting inherent broadcast advantage of wireless communication and employing a more realistic energy consumption model for wireless communication which not only depends on radio propagation losses but also on energy losses in transceiver circuitry, our simulation results show that the proposed VMB-based multicasting scheme outperforms existing prominent tree based energy conserving, path efficient multicasting schemes.

## 1. INTRODUCTION

Ad hoc networks of wireless static nodes have become a key factor in the evolution of wireless networks, especially in a rapidly deployed sensor based monitoring system. As wireless hosts in such a network are mostly driven by battery power, energy limitation is a major design constraint for this type of networks. A crucial problem in such networks is to find path(s) from source to destination(s) with minimum energy consumption. For a given source-destination pair, this problem is referred as minimum-energy unicast routing and is well investigated (for example, [1]).

Multicasting is the other prominent and useful communication mechanism for information dissemination in networks where a source node sends a message to a group of destination nodes. Broadcast is a special type of multicast communication where all nodes in the network except the source node become destination nodes. Multicast support is an important and desirable feature for wireless ad hoc networks due to its relevance in applications like large file transfer or flow of multimedia streams from one source node to several destination nodes simultaneously. In addition, many routing schemes require such communication, for updating their states and maintenance of routes between nodes. Several existing multicast routing protocols for ad hoc networks have been proposed with an aim to establish a connected distribution structure often known as *virtual multicast infrastructure* or *Virtual Multicast Backbone* (*VMB*) [2]. For a given multicast session, the entire communication process and multicast group management functions are performed only by members of VMB [2]. VMB spans or dominates all multicast group members and also includes multicast message forwarding nodes. Owing to the inherent broadcast nature of wireless channel, whenever any node transmits a message, all nodes within its transmission range receive the message if nodes use omnidirectional antennas. This property of wireless communication is also referred as *wireless multicast advantage (WBA)*. Hence, multicast routing can be simplified by arbitrarily increasing transmission power (to increase maximum range of transmission) of the source node such that all multicast destination nodes can be reached in a single transmission. But this has several disadvantages. Large transmission radius may cause decrease in effective bandwidth available per node, reduce throughput, increase interference in local neighborhood and induce faster depletion of battery power. To minimize these problems, multicast routing algorithms should select shorter distance intermediate nodes during multicast path set up. Such routes involve more number of forwarding nodes resulting in increased end-to-end delay.

 Due to scarce energy resource in ad hoc networks, minimizing energy consumption for multicasting and multicasting problems have received major attentions of researchers. They are usually referred as *minimum energy broadcasting* (MEB) and minimum energy multicasting (MEM) problems in ad hoc networks. Various solutions have been proposed for these problems by [3], [4], [5], [6], [7], [8], [9], [10]. [9] studied the multicasting problem with a relatively new objective to jointly optimize energy conservation and path efficiency. Path efficiency in this context relates to an important QoS metric, *end-to-end delay*, which is assumed to be proportional to the average number of hops between source and individual multicast destination. They proposed two tree-based multicasting algorithms to improve power savings and path efficiency. Although there have been many studies on MEM problems, there are only few works done to jointly optimize energy consumption and path efficiency for multicasting.

In this paper, we have studied the multicast routing problem in a static wireless ad hoc network to propose an energy-efficient multicast routing scheme that also improves path efficiency. Our scheme is based on a novel, efficient, distributed, scalable algorithm for VMB construction. The motivation of studying multicasting over the VMB is that nodes need not maintain any routing tables or any global topological information. They use only neighborhood information instead.

[2] observed that efficiency of any VMB based multicasting solution depends largely on its size, since it controls network overhead, energy and bandwidth consumption and end-to-end delay. So we want to generate a VMB with minimum size. The size of the VMB generated by our algorithm has an approximation ratio of 8 as compared 10 in [2], when the multicast group size is equal to the network size. It is also shown that the average size of VMB in our algorithm increases linearly with the network size for a given multicast group size while network degree is kept constant. Simulation results show that the message complexity of the proposed algorithm is almost a linear function of network size (i.e. $O(n)$). One may compare the result with the same of [2], which reports an $O(n \log n)$ message complexity. Thus, the VMB constructed by our algorithm is better than the ones constructed by the existing algorithm.

Average number of hops between source and each of the multicast destinations, using our algorithm, is much better than the same reported in [9]. In fact, our results are comparable to the average number of hops as obtained by standard shortest path algorithm. Total energy consumption by our algorithm in transmitting a data packet from a source to multicast destinations is less than the value of the corresponding metric in [9]. These facts highlight that our algorithm jointly improves energy consumption and path efficiency better than the existing algorithm.

The rest of the paper is organized as follows. In Section 2, we make brief review of the related works. Section 3 describes network model and gives some relevant definitions from graph theory.  In Section 4, we present a brief discussion on the required background, which is followed by our main VMB formation scheme. Simulation results for performance evaluation of our proposed algorithm are presented in Section 5. Finally, we conclude in Section 6.

## 2. RELATED WORK

Unfortunately, the problem of minimum energy multicasting has already been proven to be NP hard in [6].  Due to widespread interest in designing energy-efficient broadcast and multicast routing algorithms for wireless networks, several efficient heuristic algorithms were proposed in recent years.

For the minimum energy broadcasting problem, a straight greedy approach is to construct a broadcast tree that consist of the union of best unicast paths to each individual destination from the source node. This heuristic first applies Dijkstra's algorithm to obtain an SPT (Shortest Path Tree) and then to orient it as a tree rooted at the source node. Similarly, the MST (Minimum Spanning Tree) heuristic first applies Prim's algorithm to obtain an MST and then to orient it as a tree rooted at the source node. Mixed Integer Linear Programming (MILP) formulation of MEM problem, where nodes are using directional antennas, has been presented by [4]. Distributed algorithms like EWMA (Embedded Wireless Multicast Advantage) in  [6] were also proposed for the MEM and MEB problems. Their performances were found to be comparable to BIP. None of these algorithms studied minimum energy multicast tree construction problem coupled with the challenges introduced by mobility of the nodes in the network. [11] also analyzed several scalable multicasting strategies for MANET for different mobility patterns. Recently, [7] presented an efficient distributed algorithm for minimizing total RF power consumption for multicast communication in MANET. Effectiveness of the scheme was also verified in terms of energy savings and low communication overhead. Since only a small portion of the total radiated power is being captured by intended receivers, use of omnidirectional antennas is not considered suitable for energy-efficient broadcasting or multicasting in wireless ad hoc networks.  With the continuous development of switched beam and directional antennas and inefficiency of omnidirectional antennas, researchers were motivated to investigate the scope of using smart antennas for MEB or MEM problems. [8] also made a comprehensive study of MEB problem using practical directional antennas. In [12], authors present a novel zone-based distributed algorithm for Virtual Backbone formation in wireless ad hoc networks. This proposed algorithm can significantly reduce the Virtual Backbone size.

In some of the literature, the solution of MEM problem was obtained from the solution of MEB problem. The final minimum-energy multicast tree was indeed obtained by pruning from the minimum-energy broadcast tree all transmission links that are not needed to reach the members of the multicast group. For example, when applied to the multicast problem, the resulting scheme of BIP is called MIP (Multicast Incremental Power). Some algorithms, which follow a non-pruning approach, also exist, like MIDP (Multicast Incremental-Decremental Power) in [3], SPF (Shortest Path First) and MIPF (Minimum Incremental Path First) in [5]. They   perform better than MIP, but at the cost of higher complexity. [10] considered similar problems with additional assumption that each node in the network has a set of discrete levels of transmission power. [9] studied the problem of jointly optimizing power conservation and path efficiency and proposed two tree-based multicasting algorithms.  Besides, few recent schemes like [9], [13], also addressed the issue of joint optimization of energy conservation and *end-to-end delay* while constructing a multicast tree in a WANET. [13] proposed a heuristic called energy-based link replacement (ELR) algorithm for constructing an energy-efficient multicast tree for given delay constraint.

We consider *Unit-Disk Graph* ([14]) based models of ad hoc networks. [2] proved that a *minimum steiner connected dominating set* is more appropriate for *Unit-Disk Graph* based model of an ad hoc network for constructing an efficient virtual multicast backbone compared to Minimum Steiner Tree. The authors proposed a distributed algorithm for *Minimum Steiner Connected Dominating Set* with a constant approximation ratio of 10 and a message complexity of $O$ (*n log* (*n*)). To the best of our knowledge, only distributed heuristic algorithm proposed for construction of *minimum steiner connected dominating set* approximating virtual multicast backbone in wireless ad hoc network can be found in [2].

## 3. SYSTEM MODEL

The wireless ad hoc network is modeled by a connected graph $G = (V, E)$, where $V$ is the set of wireless hosts (hereinafter referred as nodes) and $E$ represents set of wireless links between pair of nodes. We make following assumptions for the ad hoc networks considered in our study. All nodes are homogeneous, i.e. maximum wireless transmission ranges are identical and all the receiver nodes have same signal detection threshold level while receiving the signal and this is normalized to one. Nodes are located on a two dimensional plane and each of them is equipped with omnidirectional antenna. Consequently, footprint of such a network will be a *Unit-Disk* graph. A link $e_{uv} \in E$ exists between nodes $u$ and $v$ only if the nodes are located within each other's wireless transmission ranges. Hence, the resulting graph is undirected. All such nodes, which are located within the maximum wireless transmission range of a node $u$ are known as its *one-hop neighbors*. In this paper, we will refer them as *neighbors* only.

Given a source node $s$ and a nonempty set $D$, $D \subseteq V$ of multicast group destinations, most of the existing schemes suggested establishment of a tree rooted at $s$ and spanning all nodes in $D$ that minimizes the total energy or power consumption for multicasting a message from $s$ to all members of $D$. But this problem can also be solved by constructing a *Steiner Connected Dominating Set* (SCDS) in an undirected graph $G = (V, E)$ representing a wireless ad hoc network. In general, a *dominating set* (DS) in a graph $G$ is a subset of $V$ which satisfies that every vertex $v \in V$ is either in this subset or adjacent to at least one vertex of this subset and a *connected dominating set* (CDS) is a dominating set that induces a connected subgraph of $G$. Given a graph $G = (V, E)$ and a subset $V_m \subset V$, a subset $S \subset V$ is called a *Steiner dominating set* of $V_m$ if every node in $V_m$ is either in $S$ or adjacent to at least one node of $S$. $S$ is called a *steiner connected dominating set* (*SCDS*) if $S$ is connected. Among all such possible SCDSs the one with minimum cardinality is known as *minimum steiner connected dominating set* (*MSCDS*). When $V_m = V$, $S$ is known as connected dominating set. Each node in a CDS is known as a dominator. A Steiner tree of $V_m$ is a subtree of G, which contains all nodes in $V_m$. Though MSCDS was known to be a better approximation compared to Steiner tree model for virtual multicast backbone (VMB) of *Unit-Disk Graph* [14] based models of ad hoc networks [2], finding an MSCDS in unit disk graphs is known to be NP-hard [2]. We want to construct a minimum Steiner connected dominating set $S$ of $\{s\} \cup D$ such that sum of energy consumption of delivering a message from source node to all multicast destination nodes over the VMB (formed by the members of this SCDS) is minimized with minimum end to end delay.

Definitions of some important terminologies, which will be used in subsequent discussion, are given below.

*Degree*: Number of 1-hop neighbors of a node.

*Parent*: Dominator of a node.

*Children:* All neighbor nodes of a dominator node $u$ ($u \in V$), which have node $u$ as their dominator, are considered as children of $u$. A non-dominator child node is referred as a *dominatee* in our discussion.

## 4. VMB Construction Algorithm

In this section, we propose an MSCDS approximation algorithm that has an improved approximation ratio when cardinality is considered and can be implemented in a fully distributed manner. Construction of virtual multicast backbone can be divided in two phases. In the first phase, we determine, an MCDS for an undirected graph $G = (V, E)$. In the second phase we choose a subset of nodes from the MCDS, which will form a MSCDS that can be used as a VMB to support a given multicast group.

### 4.1. MCDS construction

Our scheme is based on a modification of the distributed MCDS approximation algorithm proposed by [15]. Details are available in [16]. Here we make a brief review of the MCDS algorithm as proposed by [15] along with our modification. First, we drop the requirement of running distributed *leader election algorithm* and select an arbitrary node $v \in V$, which will start execution of the algorithm. This will change message complexity of overall algorithm from $O(n \log n)$ [15] to $O(n)$ [16]. We will refer this node as the leader node. The final MCDS solution will be a tree rooted at this node. We modified the algorithm by minimizing the distance between dominating nodes. The justification for this modification is to minimize RF energy consumption (which is proportional to square of the distance between dominators) of forwarding multicast message by dominating nodes. Each node in the network maintains some logical variables, which are listed in Table 1.

Table 1: List of variables any node $u$ maintains

| variable | Description |
| --- | --- |
| *id* | id of the node |
| *degree* | number of 1-hop neighbors of a node. |
| *parent* | id of the dominating node. |
| *rank* | hop distance between a node and the leader node. Rank of the leader node = 0. Rank of any other node $v$ = Rank of its parent node + 1. |
| *state* | state of the node. |
| *neighbor_list* | a list of neighbors of a node |
| *distance_list* | list of distances of all neighbors of a node |
| *children_list* | ids of all children nodes of a node |
| *cost* | cost of the node |
| *multicast_forwarding_status* | Whether acts as a forwarding node for source or destination or both or none |
| *forwarding_id* | Stores id(s) of the node(s) for which it acts as a forwarding node |

During the execution of the algorithm, nodes exchange following three messages as listed in Table 2, to update some of variables mentioned in Table 1. The variable *distance_list* is populated during the initial exchange of '*hello*' messages among nodes in the network. The *cost* variable stores distance of the node from its dominator. This variable is initialized with a very high value; say ten times of the maximum wireless transmission range of the node. While running the algorithm, at any time, a node *u* can be in one of the four possible states: *ordinary*, *dominator*, *dominatee* and *active*. At the beginning of the algorithm, all hosts are assumed to be in *ordinary* state. A dominatee node must have at least one neighbor which is in dominator state. Similarly, an active node (i.e. a node in *active* state) must have at least one dominatee as its neighbor. An active node is a candidate dominator in next pass of the algorithm. An active node becomes a dominator node if its cost is smallest compared to all of its active neighbors. Node *id*s will be used in case of a tie. Below, we review the transition steps [16]. Necessary state transition diagram is shown in Figure 1.

Table 2: List of messages exchanged during execution of the MCDS algorithm

| messages | parameters | description |
|---|---|---|
| *DOMINATOR* | node-id, parent-id, rank | A node *u* (*u*∈ *V*) broadcasts this message to all neighbors when it becomes a dominator. If a node receives more than one < DOMINATOR > message it chooses node with minimum rank as its dominator. |
| *DOMINATEE* | node-id, parent-id, rank | Node *u* (*u*∈ *V*) broadcasts this message to all neighbors when it becomes a dominate. When dominator node with id equals parent-id receives this message, it includes id of node *u* in its *children_list*. |
| *ACTIVE* | node-id | When a node *u* becomes *active* it broadcasts this message to all the neighbors, which are neither a dominator nor a dominatee. |

Execution of the algorithm is started with the leader node.

1. If *u* is the leader node, *u* becomes a *dominator* and sets its variables *parent=u*, and *rank*=0. *u* will broadcast message <*dominator*(*u*, *parent*, *rank*)>.

2. When an *ordinary* node *u* receives message <*dominator* (*v*, *p*, *k*)>, it enters a into *dominatee* state. This is also true for a node in *active* state. If node *p* is a neighbor of *u*, then *u* sets its variables *parent=p*, *rank=k*, and updates *cost*=distance (*p*, *u*); otherwise, *parent=v*, *rank=k*+1, *cost* =distance (*v*, *u*). *u* will broadcast message <*dominatee* (*u*, *parent*, *rank*)>.

3. If *u* is in *ordinary* state and receives message <*dominatee* (*v*, *p*, *k*)> from neighbor *v*, it goes to *active* state. Its sets variables *parent=v* and *rank=k*+1 and updates *cost*=distance (*v*, *u*). Though we have defined the variables *parent* and *cost* to store id of the dominator and distance from the dominator respectively, here *v* is not a dominator. However, at the end of the execution of the algorithm any node will be in either of two states; (i) *dominator* or (ii) *dominatee*. Hence, this is a temporary setting. All ordinary and active nodes maintain an additional *active_node_list*, which is a list of *active* node *id*s. If *v* is in *active_node_list*, it will be removed. *u* will broadcast message <*active* (*u*, *cost*)>.

4. If *u* is in *active* state and receives message *<dominatee (v, p, k)>* from dominatee *v*, it checks if *cost* of node *u* with dominator *v* is less than existing cost of *u*. If that is indeed the case, then *parent = v, rank = k*+1 and *cost*=distance (*u, v*). *u* remains in active state. If *v* is in *active_node_list*, it will be removed.

5. If *u* is in ordinary or in active state and receives a message *<active (v, cost)>* state of *u* remains unchanged and *v* is included in *active_node_list*.

6. If an active node *u* finds its *cost* to be smallest among all nodes in its *active_node_list*, and if there is no broadcasting in its neighborhood for predefined constant time period *t*, then *u* becomes a dominator. It will broadcast a message *<dominator (u, p, k)>*.

7. If a dominator node *u* finds none of its neighbors considers *u* as its dominator, *u* changes its state to *dominatee*. *u* will broadcast a message *<dominatee (u, parent, rank)>*.

8. A *dominatee* node *u* becomes a *dominator* node if it receives the message *<dominatee (v, u, k)>* from dominator *v*.

At the end of the first phase, all hosts, which are in *dominator* state, form the minimum connected dominating set. After completion of the algorithm for MCDS, rank of each node in the network indicates its distance (in number of hops) from the leader node through intermediate dominator nodes. We exploit this information in the second stage of the algorithm. When multicast group size equals the network size (broadcast case), the MCDS obtained by above mentioned stage could be utilized as a VMB for the network. If the size of the multicast group is smaller than network size then the necessary steps for VMB formation are described below.

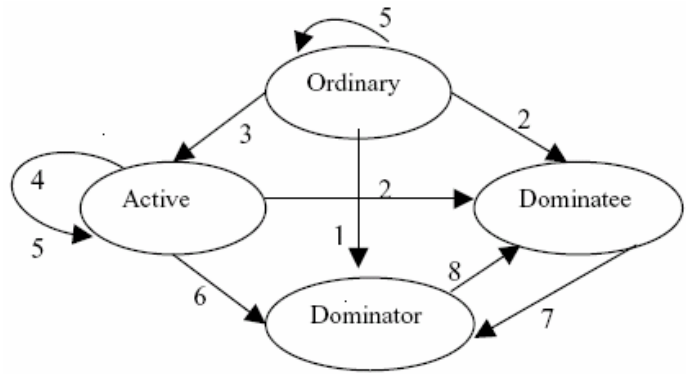

Figure1: the state transition diagram for MCDS formation for any node *u*

## 4.2. VMB formation

We assume that the source node knows *ids* of all multicast destination nodes. When a source node wants to initiate a multicast session, first, it verifies whether all multicast destination nodes are its neighbors by checking its *neighbor_list*. If so, computation of the VMB is needless, since all destination nodes can be reached with a single transmission. But if the source node finds one or more destinations are not in the *neighbor_list*, it sends a message *<VMB_SOURCE_REQUEST>* to its parent. The message includes ids of source and destination nodes in the multicast group. On receiving this message, parent of source node (say *v*) fixes its status (updates the variable *multicast_forwarding_status*) as *source_forwarding_node*. Then it forwards *<VMB_SOURCE_REQUEST>* to its parent. Parent of *v* performs same functions as *v*. This will continue until the leader node receives message *<VMB_SOURCE_REQUEST>*. When leader node receives *<VMB_SOURCE_REQUEST>* message, it also becomes (updates *multicast_forwarding_status* ) *source_forwarding_node*. In this way, a path is set up from source node to the leader node through dominating nodes. All dominating nodes with

*source_forwarding_node* status form *source_forwarding_nodes_set*. It is assumed that each node knows for which multicast sessions it is a member of the destination set. Each multicast destination node sends a message <*VMB_DESTINATION_REQUEST*> to its parent. This message includes *id* of the destination node that initiates the message. The parent node updates its status to *destination_forwarding_node* and stores *id* of the destination node in *forwarding_node_id*. Then, it forwards <*VMB_DESTINATION_REQUEST*> to its parent. This process will continue until the message reaches the leader node. Thus, each of the multicast destination nodes independently sets up a path, through leader node, to reach the source node using a subset of dominating nodes. All dominating nodes with *destination_forwarding_node* status together form *destination_forwarding_nodes_set*. If only the leader node is selected as *destination forwarding node* for all multicast destinations, then final solution for VMB is union of *source_forwarding_nodes_set* and *destination_forwarding_nodes_set*. Otherwise, there exists one or more dominator nodes (other than leader node), whose, *forwarding_node_id* stores *id*s of each member of the multicast group. If any such node receives a message from source node, it can send it to all multicast destinations by forwarding it to other members of *destination_forwarding_nodes_set*. Now, if more than one such node exists, then we have to identify the node with highest rank (say, *v* be the node). Hence, all dominating nodes with *source_forwarding_node* status and having a rank lesser than that of *v* will be redundant for the final solution. Final solution consists of union of *destination_forwarding_nodes_set* and subset of *source_forwarding_nodes_set* having rank greater than or equal to that of *v*.

Let us illustrate the scheme with the help of the following example. The given unit-disk graph *G* = (*V*, *E*) is shown in Figure 2. There are twelve nodes with *ids* from 1-12. Node 1 is arbitrarily chosen to be the leader node. Final MCDS obtained applying the distributed implementation of our modified algorithm is {1, 2, 3, 8 and 12}. Parent node for each node is given in Table 3. Let us assume that in a given multicast session, node 5 is the source node (indicated by a triangle in Figure 2) and destination nodes are 4, 9 and 12 (indicated by diamonds in Figure 2). Different steps of the VMB construction algorithm as applied to the given example are described in the following.

1. At the beginning, source node 5 finds only one multicast destination node (node 9) in its neighborhood, so it decides to start the second phase of the algorithm. Since node 8 is the parent of node 5, it sends a message <*VMB_SOURCE_REQUEST*> to node 8. After receiving the message node 8 fixes its status as *source_forwarding_node*. The node also forwards message <*VMB_SOURCE_REQUEST*> to its parent node 1. Now, node 1 fixes its status as *source_forwarding_node*. Thus a path from source node to leader node is established in a distributed way and the path is {5, 8, 1}.

2. Node 4 now performs following steps to set up a path to reach the leader node 1. It sends a message <*VMB_DESTINATION_REQUEST*> to its parent node 2. Node 2 fixes its status as *destination_forwarding_node* and forwards a message <*VMB_DESTINATION_REQUEST*> to its parent node 1. On receipt of this message, node 1 fixes its status as *destination_forwarding_node*. Thus, a path is established between node 4 and leader node 1 and the same is {4, 2, 1}.

3. Similarly, node 12 sets up a path to node 1. The path is {12, 3, 8, 1}.

Final VMB obtained to support the multicast session is union of all source and destination forwarding nodes. Therefore, the final solution is {1, 2, 3, 8} which is a subset of the MCDS {1, 2, 3, 8 and 12}.

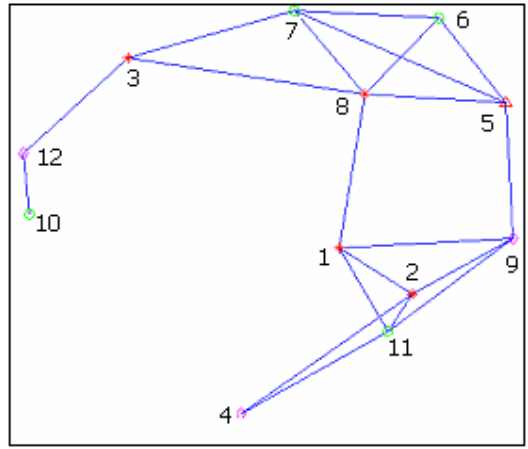

Figure 2: An example unit-disk graph *G* contains 12 nodes and 19 links

Table 3: parent(dominator) node for each node in Figure 2

| Node id | 1 | 2 | 3 | 4 | 5 | 6 | 7 | 8 | 9 | 10 | 11 | 12 |
|---|---|---|---|---|---|---|---|---|---|---|---|---|
| Parent node id | 1 | 1 | 8 | 2 | 8 | 8 | 8 | 1 | 1 | 12 | 1 | 3 |

***Lemma 1***: *Size of the VMB generated by our algorithm is at most 8×opt + 1, where opt is the size of the minimum connected dominating set.*

***Proof***: In our scheme, all VMB members are essentially members of the MCDS obtained by running *MCDS construction* phase of our algorithm. Hence, they are all dominators. When multicast group size is less than number of nodes in the network, our VMB will be a subset of the MCDS of the graph corresponding to the network. When multicast group size equals network size our VMB equals MCDS. Size of the MCDS generated by the algorithm (Cheng et al. 2006) is at most 8×*opt* + 1, where *opt* is the size of the minimum connected dominating set. Thus, the VMB obtained by our scheme has maximum size of 8×*opt* + 1 which is better than the same obtained by (Ya-feng et al. 2004). Besides, another important feature of our solution is that nodes need not maintain any routing table that stores information about distance and next hop neighbors' *ids* as required in the VMB formation scheme discussed in (Ya-feng et al. 2004).

***Lemma 2:*** *Message complexity of our distributed algorithm is O (mn), where n is the number of nodes in the network and m is the multicast group size.*

***Proof***: The MCDS approximation algorithm considered here has a message complexity *O (n)*, where *n* is the number of nodes in the network (Cheng et al. 2006, Cheng and Du, 2002). For the second phase of our algorithm, total number of messages required to be transmitted for discovery of a path from any member of the multicast group to the leader node is at most *O (n)*. Hence, message complexity for the second phase of our algorithm is *O (mn)*. Final VMB will be union of all intermediate nodes of such paths along with the leader node. Hence total message complexity is *O (n)* +*O (mn)*= *O (mn)*.

Since a VMB is constructed to support a given multicast session, we need as many VMBs as the number of multicast sessions in the network. In case of constructing multiple VMBs for a given network we need to execute the first phase (i.e. MCDS construction) of our algorithm only once.

VMB formation phase needs to be executed once for each VMB construction. This will also improve average message complexity for constructing multiple VMBs to support multiple multicast sessions.

### 4.3. Multicast routing over VMB

Routing of a multicast message over a given VMB is explained in the following.

1. A VMB is used to support a given multicast session. In a given multicast session, whenever source node wants to send a message to destination nodes it first searches for destination nodes and VMB member nodes. Following the concept of SCDS, the source node and each of the multicast destination nodes will have at least one VMB neighbor. Now source sends the message to the nearest *source_forwarding_node* and one more multicast destination if any, in its neighborhood. The power requirement is estimated considering "wireless multicast advantage". Each VMB neighbor will forward the message to downstream VMB neighbors (which have not received the message yet). In other words, the message will be flooded in the VMB. This will continue until parent nodes of all multicast destination nodes receive the message. Then, in the next transmission session these VMB members will deliver the message to all multicast destination nodes. Since multicast messages will be flooded only through VMB members, the rate of redundant transmission in the network is significantly reduced and broadcast storm (Ni et al. 1999) problem can be substantially minimized.

2. Let us consider the opposite scenario when a destination node wants to send any control message to the source node. It only needs to send the message to its parent node (which must be a VMB member). The parent node will forward the message to its parent (also a VMB member). This will continue until a *source_forwarding_node* gets the message. When a *source_forwarding_node* gets the message, it forwards the message to its upstream *source_forwarding_node* (which has not received the message yet). This process will continue until parent of the source node receives the message. In the next transmission session, parent of source node sends the message to source node.

Note that, in our proposed scheme we do not require route discovery or routing table information exchange for finding and maintaining routes from the source to each multicast destination node. Thus, we significantly reduce the routing overhead, which is a major challenge in view of the scarce communication resources and dynamic nature of the underlying network topology of wireless ad hoc networks.

## 5. PERFORMANCE EVALUATION

In this section, we present results of simulation experiments to evaluate performance of our VMB based multicasting scheme. Performance metrics considered are: (i) average size of VMB, (ii) message complexity of VMB construction process, (iii) total power consumption in transmitting a data packet from source to all multicast destinations, and (iv) average number of hops from source node to each multicast destination. The first two metrics reflect the scalability of the VMB construction algorithm. The last two metrics reflect energy conservation and path efficiency of the proposed multicasting scheme.

Our simulation models a static wireless ad hoc network by two-dimensional random graphs while the lower MAC layer is assumed to be ideal. Random graphs are generated in $10m \times 10m$ area by randomly throwing $n$ nodes. For a given wireless transmission range $r$, an edge is added between each pair of nodes that has a Euclidian distance less than or equal to $r$. One of the nodes is randomly chosen as source node. Multicast destinations nodes are also chosen randomly from the remaining nodes in the network depending upon the size of the multicast group.

In order to gauge the scalability of our scheme, we have considered scenarios with a specified network size $n$ ($n$=40, 80, 120, 160, 200), a multicast group size $m$ ($m$=25%, 50%, 75% and 100% of the network size) and a constant average node degree $\overline{\lambda}$ ( $\overline{\lambda}$=5, 10 and 15). If the area of the graph and $r$ are fixed, then $\overline{\lambda}$ increases with $n$. Therefore, to maintain a constant $\overline{\lambda}$, $r$ is adjusted as $n$ increases. In all cases, i.e., for any combination of network size $n$, multicast group size $m$ and average degree $\overline{\lambda}$, we randomly generated 50 different instances and results shown in Figure 3 to Figure 6 are averages over 50 such instances. These figures show the variation of average VMB size as the function of network size with average node degree as a parameter. As the figures show, average size of VMB increases almost linearly. This satisfies our initial goal that the proposed solution for VMB should be scalable.

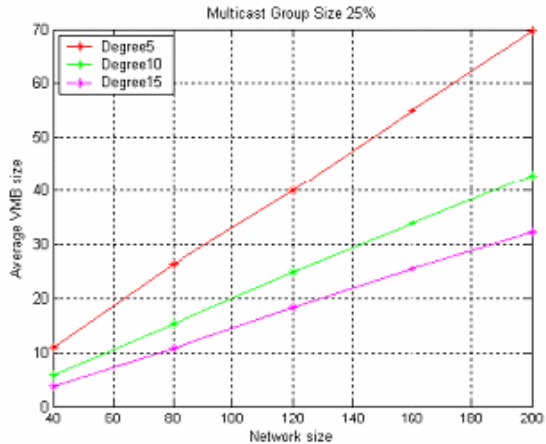

Figure 3: Average VMB size as a function of network size and average degree for $m$=25%

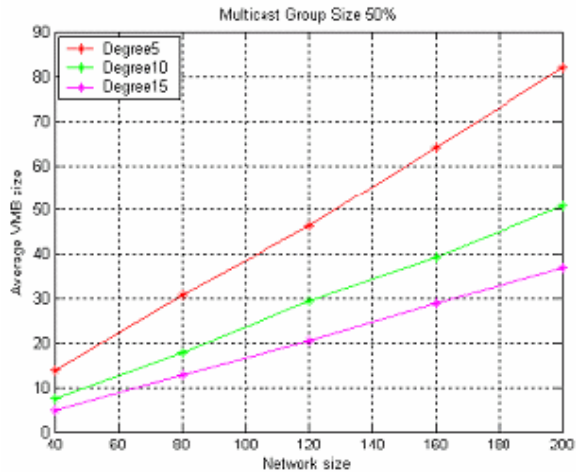

Figure 4: Average VMB size as a function of network size and average for $m$=50%

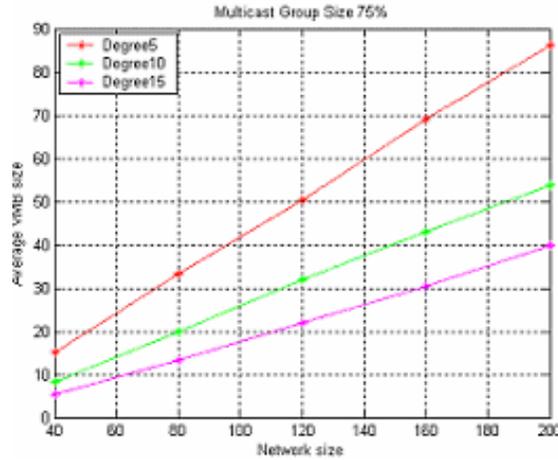

Figure 5: Average VMB size as a function of network size and average degree for *m*=75%

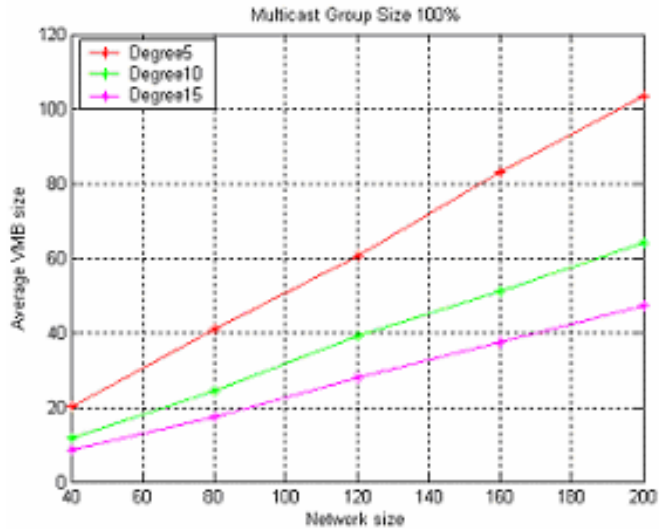

Figure 6: Average VMB size as a function of network size and average degree for *m*=100%

Now, we evaluate performance of the proposed algorithm in terms of the communication overhead. In Figure 7, we show variation of *normalized communication overhead* with change in network size. Normalized communication overhead is defined as the ratio of total number of control messages needed to be exchanged for VMB formation to the network size. Simulation parameters for this set of experiments are as follows. Simulation area is 10m × 10m, maximum wireless range *r* of nodes is set to be 2.5m. Number of nodes is varied form 40 to 400. Multicast group size is considered as a parameter. Three different values of this parameter considered here are 25%, 50% and 100% of overall size. We randomly generated 50 different instances and we present here the average over those 50 instances. Figure 7 clearly shows that the total number of messages exchanged increases almost linearly with increase in network size under a fixed multicast group size. It also indicates that communication overhead increases with increase in multicast group size. Both observations verify our analytical results on message complexity. In fact, simulation results indicate that there exists a much stronger bound of message complexity.

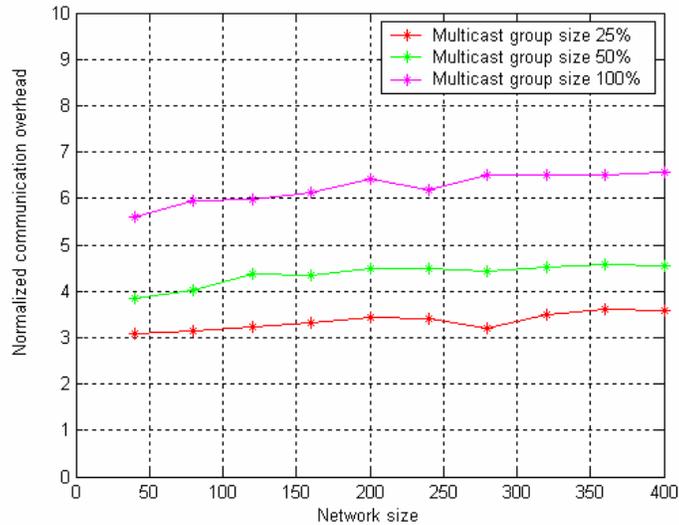

Figure 7: Normalized communication overhead as a function of network size and multicast group size

For the third performance metric, we consider the following energy consumption model [17].

$$E_{tx}(d) = E_{elec} + Kd^{\alpha} \ldots\ldots\ldots\ldots(1)$$

$$E_{total} = E_{tx}(d_{ij}) + E_{rx} \ldots\ldots\ldots(2)$$

where $E_{tx}(d)$ is the energy required for transmitting a data packet over a distance $d$. $E_{total}$ is the total energy (including energy consumption at receiver) required for sending a data packet over a distance $d$. . $E_{elec}$ represents energy consumption for transmission processing (modulation, encoding etc. ) to transmit a data packet. $E_{rx}$ denotes the energy cost associate with reception processing. $E_{rx}$ is assumed to be same for every node. $\alpha$ is the path loss constant (in this paper we have assumed $\alpha$=2) and K is another constant (here K=1). For short range radio communication $E_{elec} = 4r^{\alpha}$ and $E_{rx}$=0.7*$(E_{elec}+Kd^{\alpha})$ [17]. We ignore the energy consumption while a node remains simply "on" without transmitting or receiving. If a transmitting node has more than one receiving neighbors, due to *WBA* property considered here, transmitting energy would be proportional to maximum distance between transmitting node and receiving nodes. In such case, total energy consumption will be

$$E_{total} = E_{tx}(d_{max}) + pE_{rx} \ldots\ldots\ldots(3)$$

where, $p$ is the number of receiving nodes and $d_{max}$ is the maximum distance between transmitting node and receiving nodes.

Number of nodes considered for this set of experiments is $n$=50 and 100. Maximum transmission range is 2.5m and area remains 10m × 10m. We compare the performance of our algorithm with two other algorithms mentioned in (Ye et al. 2004). Three algorithms considered for comparison are as follows: (i) our proposed algorithm (VMB), (ii) *Minimum Spanning Tree Heuristic* (MSTH) (Ye et al. 2004), which is based on the multicast incremental power (MIP) algorithm proposed in (Ya-feng et al. 2004), (iii) MSTH-II (Ye et al. 2004), a variation of MSTH algorithm that jointly optimizes power conservation and path efficiency. Figures 8 and 9 show results of total energy consumption of multicast trees, obtained using three different multicasting algorithms, for different size of the multicast group.

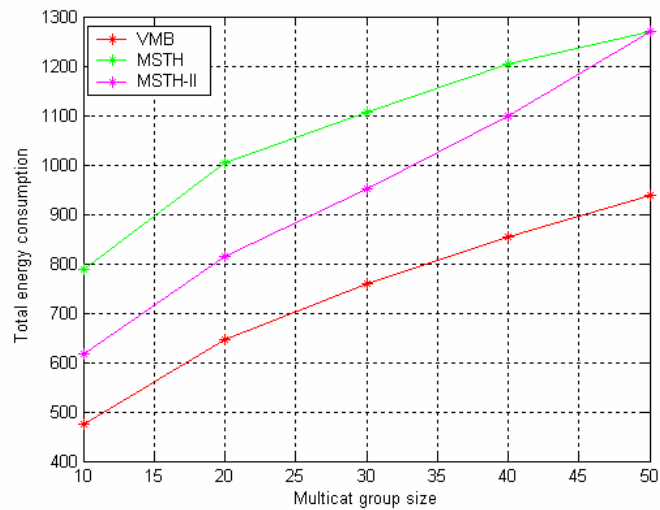

Figure 8: Performance comparison for total energy consumption for *n*=50, other parameters are r=2.5, K=1.

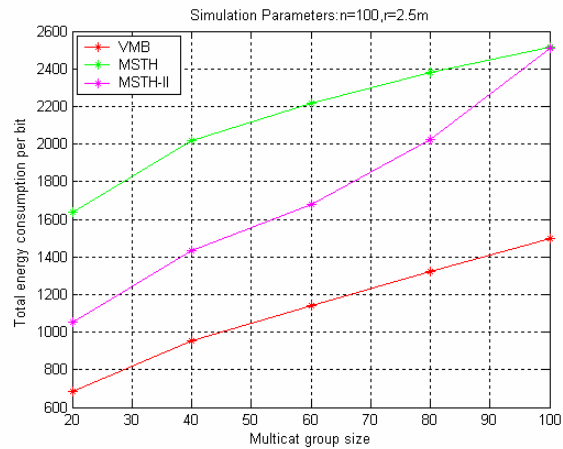

Figure 9: Performance comparison for total energy consumption for *n*=100, other parameters are r=2.5, K=1.

In Figures 8 and 9, simulation results clearly indicate that our proposed scheme yields better energy conservation compared to MSTH and MSTH-II. The reason being in (Ye et al. 2004) the energy consumption by the receiving nodes and energy consumption in transceiver circuitry of the transmitting nodes were completely ignored. Instead, power consumption was considered to be entirely dependent on the Euclidean distance between transmitting and receiving nodes. Simulation results in [9] might not reflect actual energy consumption scenario in for short-range radio communication in ad hoc wireless networks. In order to conserve power, algorithms of [9] avoided long distance routes and preferred low power hops. This might lead to increased multicast tree size and higher energy consumption.

In the fourth and last set of experiments, we compare path efficiency of these three candidate algorithms by computing average hop count between source node and individual multicast destinations. In addition to these three algorithms, we have implemented Unicast Shortest Path (USP) algorithm, which serves as a limit of the maximum achievable path efficiency. USP is the conventional per-source shortest path algorithm using hop count as the routing metric. Results shown in Figure 10 and Figure 11 indicate that our VMB based multicasting algorithm clearly outperforms MSTH and MSTH-II and performs competitively with USP.

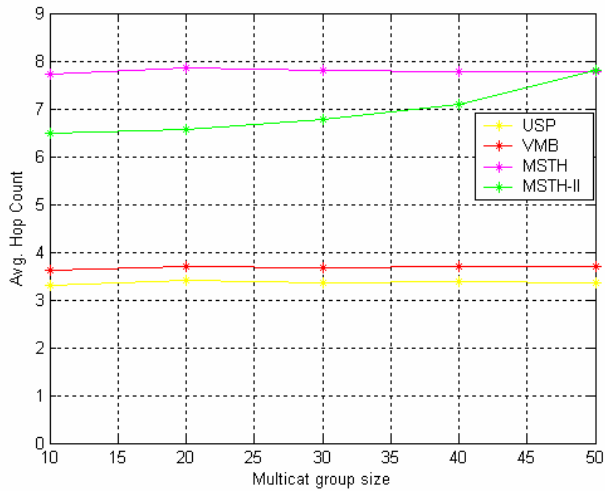

Figure 10: Performance comparison for average hop count for *n*=50.

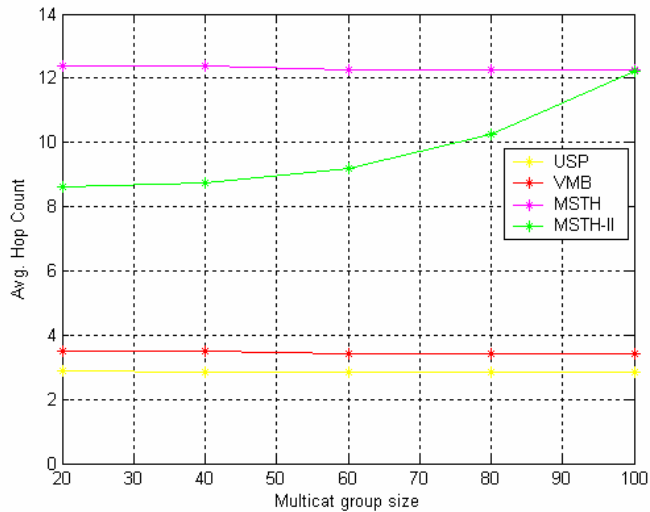

Figure 11: Performance comparison for average hop count for *n*=100.

## 6. CONCLUSION

We have presented a simple distributed, scalable, energy-efficient, heuristic algorithm for constructing a virtual multicast backbone with smaller number of forwarding nodes in ad hoc network. MSCDS in Unit Disk Graph is used to model the VMB. Formation of this VMB does not require maintenance of routing tables at each node, which could be formidable task in view of the dynamic nature of the underlying topology of the ad hoc network. We have also proposed a simple multicast routing based on this VMB where nodes need not maintain any routing table. Simulation results demonstrate that our scheme achieves significant improvement in path efficiency and also succeeds in energy conservation compared to other power efficient multicast routing strategies.

**Author's identification**

1.  Name: Tamaghna Acharya
    Mailing Address : Dept. of ETCE, BESU, Shibpur, Howrah-711103, India
    e-mail: t_acharya@telecom.becs.ac.in

Brief biographical sketch: Tamaghna Acharya received his Bachelor's degree in Electronics and Telecommunication Engineering from Bengal Engineering College (Deemed University), Shibpur, India in 2000. He received his M.Tech. degree in telecommunication systems engineering from Electronics and Electrical Communication Engineering Dept. of IIT Kharagpur in 2005. Presently, he is a lecturer in the Electronics & Telecommunication Engg. Dept. of Bengal Engineering & Science University Shibpur, India.

Photograph:

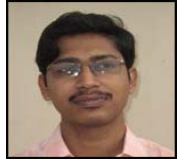

2.  Name: Samiran Chattopadhyay

Mailing Address : Dept. of Information Technology, Jadavpur University, Kolkata-700098, India
e-mail: samiranc@it.jusl.ac.in

Brief biographical sketch: Samiran Chattopadhyay received his Bachelor of Technology (B.Tech) degree in Computer Science and Engineering from IIT Kharagpur in 1987. He received his M.Tech. degree from the same department in 1989 and received his Ph.D Degree from Jadavpur University in 1993. Presently, he is a professor in the department of Information Technology, Jadavpur University, India.

Photograph:

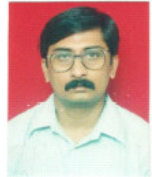

3.  Name: Rajarshi Roy
    Mailing Address : Dept. of Electronics and Electrical Communication Engineering, Indian Institute of Technology, Kharagpur, India
    e-mail: royr@ece.iitkgp.ernet.in

Brief biographical sketch: Rajarshi Roy received the Bachelor's degree in Electronics and Telecommunication Engineering from Jadavpur University, Kolkata, India, with first class honors in 1992 and the MSc (Engg) degree in Electrical Communication Engineering from the Indian Institute of Science, Bangalore, Karnataka, India, in 1995. He received the PhD degree in Electrical Engineering with a specialization in Networking and a Minor in Computer Science from Polytechnic University, Brooklyn, New York (Now known as Polytechnic Institute of New York University, New York University, New York, NY, USA) in 2001. Since July 2002, he has been employed as an Assistant Professor in the Department of Electronics and Electrical Communication Engineering, Indian Institute of Technology, Kharagpur, West Bengal, India. He has worked for Comverse, MA, USA and Lucent, Bangalore, India. He served Helsinki University of Technology and Indian Statistical Institute, Kolkata as academic visitor and served Bell Labs, Holmdel, NJ, USA as summer intern. His research interest includes communication networks, queuing theory, optimization, algorithms and resource allocation problems.

Photograph:

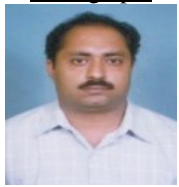